\documentstyle[twocolumn,eqsecnum,prl,aps]{revtex}

\def\sqr#1#2{{\vcenter{\hrule height.#2pt\hbox{\vrule width.#2pt
height#1pt \kern#1pt \vrule width.#2pt}\hrule height.#2pt}}}

\draft
\newcommand{\be}{\begin{equation}}
\newcommand{\ee}{\end{equation}}
\newcommand{\ben}{\begin{eqnarray}}
\newcommand{\een}{\end{eqnarray}}

\begin{document}
\draft

\twocolumn[\hsize\textwidth\columnwidth\hsize\csname
@twocolumnfalse\endcsname

\title{Naudts-like duality and the extreme Fisher information principle}
\author{L. P. ~Chimento$^{1,\,3}$, ~F. ~Pennini$^2$,  and A.
Plastino$^{2,\,3,*}$}
\address{$^1$Departamento de F\'{\i}sica, Facultad de Ciencias
Exactas y Naturales, \\ Universidad de Buenos Aires, Ciudad
Universitaria, Pabell\'{o}n I, 1428 Buenos Aires, Argentina.}
\address{$^2$Departamento de F\'{\i}sica, Universidad Nacional de La Plata,
C.C. 727, 1900 La PLata, Argentina.\\ $^3$ Argentine National
Research Council (CONICET)\\ $^*$
E-mail:~plastino@venus.fisica.unlp.edu.ar}
 \maketitle

\begin{abstract}
We show that using  Frieden and Soffer's extreme information principle (EPI)
\cite{f7} with a Fisher measure constructed with escort probabilities
\cite{beck}, the concomitant solutions obey a type of Naudts' duality
\cite{naudts} for nonextensive ensembles \cite {review}. \pacs{ 02.50.-r,
89.70.+c, 02.50.Wp, 05.30.Ch} KEYWORDS: Fisher information, extreme physical
information, Tsallis entropy, escort probabilities.\vspace{0.5 cm}
\end{abstract}]


\section{\strut Introduction}


We are going to be concerned in what follows with the workings of two
information measures that have received much attention lately, those of Fisher
\cite{f7,roybook} and Tsallis \cite{review,t0,t00}, respectively. Our goal is
to show that their interplay naturally yields a type of Naudts' duality
\cite{naudts}.

Fisher's information measure (FIM) \cite{f7,roybook} was advanced already in
the twenties, well before the advent of Information Theory (IT), being
conventionally designed with the symbol $I$ \cite{roybook} (see Eq. (\ref
{definir}) below for the pertinent definition). Much interesting work has been
devoted to the physical applications of FIM in recent times (see, for instance,
\cite {f7,roybook,pla2,pla4} and references therein). Frieden and Soffer
\cite{f7} have shown that Fisher's information measure provides one with a
powerful variational principle, the extreme physical information (EPI) one,
that yields the canonical Lagrangians of theoretical physics \cite{f7,roybook}.
Additionally, $I$ has been shown to provide an interesting characterization of
the ``arrow of time'' alternative to the one associated with Shannon's $S$
\cite{pla2,pla4,pla5}.

Tsallis' measure is a generalization of Shannon's one. Notice that IT was
created by Shannon in the forties \cite{jaynes,katz}. One of its fundamental
tenets is that of assigning an information content (Shannon's measure) to any
normalized probability distribution. The whole of statistical mechanics can be
elegantly re-formulated by extremization of this measure, subject to the
constraints imposed by the {\it a priori} information one may possess
concerning the system of interest \cite{katz}. It is shown in \cite
{review,t0,t00} that a parallel process can be undertaken with reference to
Tsallis' one, giving rise to what is called Tsallis' thermostatistics,
responsible for the successful description of an ample variety of phenomena
that cannot be explained by appeal to the conventional one (that of
Boltzmann-Gibbs-Shannon) \cite{review,t0,t00}.

\section{A brief Fisher primer}

Fisher's information measure $I$ is of the form
\begin{equation}  \label{definir}
I\,=\,\int \,dx\,f(x,\theta )\,\left[ \frac{1}{f(x,\theta )}\,\frac{\partial
f}{\partial \theta }\right] ^{2},
\end{equation}
where $x$ is a stochastic variable and $\theta$ a parameter on
which the probability distribution $f(x,\theta)$ depends. The
Fisher information measure provides a lower bound for the
mean-square error associated with the estimation of the parameter
$\theta$. No matter what specific procedure we chose in order to
determine it, the associated mean square error $e^2$ has to be
larger or equal than the inverse of the Fisher measure
\cite{roybook}. This result, i.e., $e^{2}\,\ge \,\frac{1}{I}, $ is
referred to as the Cramer-Rao bound, and constitutes a very
powerful statistical result \cite{roybook}.

The special case of {\it translation families} deserves special mention.
These are mono parametric families of distributions of the form $f(x-\theta
) $ which are known up to the shift parameter $\theta $. Following Mach's
principle, all members of the family possess identical shape (there are no
absolute origins), and here Fisher's information measure adopts the
appearance
\begin{equation}  \label{tra}
I=\int dx \, \frac{1}{f} \,\left[\frac{\partial f}{\partial x}\right]^2.
\end{equation}
The parameter $\theta$ has dropped out. $I=I[f]$ becomes then a functional of
$f$.

At this point we introduce the useful concept of escort probabilities (see
\cite{beck} and references therein), that one defines in the fashion

\begin{equation}
F_q(x)=\frac{f(x)^{q}}{\int f(x)^{q}dx},  \label{escort}
\end{equation}
$q$ being any real parameter, $\int F_q(x) dx=1$, and, of course, for $q=1$
we have $F_1\equiv f$.

The concomitant ``escort-FIM" becomes
\begin{equation}  \label{Iq}
I[F_q]\,=\int \,dx\,F_q(x)\,\left[ \frac{1}{F_q(x)}\, \frac{\partial F_q(x)}{%
\partial x}\right] ^{2},
\end{equation}
that, in terms of the original $f(x)$ acquires the aspect
\begin{equation}
I[F_q]=q^{2}\frac{\int \,dx\,f(x)^{q-2}\left[ \frac{\partial f(x)}{\partial x%
}\right] ^{2}}{\int dx\,f(x)^{q}}.  \label{Iescort}
\end{equation}

We shall denote with $I_{q}$ the new ``escort-FIM''
\begin{equation}
I_{q}=\frac{\int \,dx\,f(x)^{q-2}\left[ \frac{\partial f(x)}{\partial x}%
\right] ^{2}}{\int dx\,f(x)^{q}}.  \label{Iescortq}
\end{equation}
(Notice that for $q=0$ the integration range must be finite in order to avoid
divergences in the denominator.)

The parameter $q$ can be identified with Tsallis' nonextensivity index \cite
{t2,t3,t4}, which allows one to speak of ``Fisher measures in a nonextensive
context''. Their main properties have been discussed in \cite {Renyi}.

\section{The extreme physical information principle (EPI)}


The Principle of Extreme Physical Information (EPI) is an overall physical
theory that is able to unify several sub-disciplines of Physics \cite
{f7,roybook}. In Ref. \cite{f7} Frieden and Soffer (FS) show that the
Lagrangians in Physics arise out of a mathematical game between an
intelligent observer and Nature (that FS personalize in the appealing figure
of a ``demon", reminiscent of the celebrated Maxwell's one). The game's
payoff introduces the EPI variational principle, which determines
simultaneously the Lagrangian {\it and} the physical ingredients of the
concomitant scenario.

FS \cite{f7} envision the following situation, involving Fisher's
information for translation families: some physical phenomenon is being
investigated so as to gather suitable, pertinent data. Measurements must be
performed. Any measurement of physical parameters appropriate to the task at
hand initiates a relay of information $I$ (or $I_q$ in a non-extensive
environment) from Nature (the demon) into the data. The observer acquires
information, in this fashion, that is precisely $I$ (or $I_q$). FS assume
that this information can be elicited via a pertinent experiment. Nature's
information is called, say, $J$ \cite{f7,roybook}.

Assume now that, due to the measuring process, the system is perturbed,
which in turn induces a change $\delta J$. It is natural to ask ourselves
how the data information $I_q$ will be affected. Enters here FS's EPI: {\it %
in its relay from the phenomenon to the data no loss of information should
take place}. The ensuing new Conservation Law states that $\delta J=\delta
I_q$, or, rephrasing it
\begin{equation}
\delta (I_q\,-\,J)\,=\,0,  \label{epiq}
\end{equation}
so that, defining an action ${\cal A}_q$
\begin{equation}
{\cal A}_q\,=\,I_q\,-\,J,  \label{action}
\end{equation}
EPI asserts that the whole process described above extremizes ${\cal A}_q$.
FS \cite{f7,roybook} conclude that the Lagrangian for a given physical
environment is not just an {\it ad-hoc} construct that yields a suitable
differential equation. It possesses an intrinsic meaning. Its integral
represents the physical information ${\cal A}_q$ for the physical scenario.
On such a basis some of the most important equations of Physics can be
derived for $q=1$ \cite{f7,roybook}. For an interesting Quantum Mechanical
derivation see \cite{Penni1}. A cosmological application of the nonextensive
($q \ne 1$) conservation law (\ref{epiq}) is reported in \cite{chifla}.
Mechanical analogs that can be built up using this law are discussed in \cite
{fla}. Notice, however, that the last two references use an old Tsallis'
normalization procedure (advanced in \cite{t2,t3}), that cannot be
assimilated within the framework of the escort distribution concept.

\section{Solutions to the variational problem}

According to EPI, $J$ is fixed by the physical scenario \cite{roybook}. We
adopt here a more modest posture by assuming that $J$ embodies only the
normalization constraint, and say nothing regarding a specific physical
scenario.  $%
J $ is just

\begin{equation}  \label{Jq}
J\,=\lambda\,\int \ f(x)\ dx,
\end{equation}

\noindent where $\lambda $ is the pertinent Lagrange multiplier. Such a $J$
has been successfully employed in \cite{Penni1} with reference to a quantum
mechanical problem. Playing the Frieden-Soffer game, i.e., performing the
variation (\ref{epiq}), leads then to

\begin{equation}
2f\ \ddot{f}+(q-2)\ \dot{f}^2+q\ I_q\ f^2+\lambda \ Q\ f^{3-q}=0  \label{f}
\end{equation}

\noindent a $q$-dependent, non-linear differential equation that should
yield our ``optimal'' probability distribution $f$ (we set $Q=\int f^{q}dx$%
). Now, one should demand that, for $q=1$, (\ref{f}) become identical to the
differential equation that arises in such circumstances (see that equation
in \cite{Penni1}, for instance, and call $\lambda ^{^{\prime }}$ the
concomitant Lagrange multiplier used there). This requirement is fulfilled
if we set $\lambda =\lambda ^{^{\prime }}-qI_{q}$. The $q=1$-expression
becomes then
\begin{equation}
2f\ \ddot{f}-\ \dot{f}^{2}+\lambda ^{\prime }\ f^{2}=0,  \label{q1}
\end{equation}
where, of course, one has $Q=1$. The solution of Eq.
(\ref{q1}) is of the form

\begin{equation}
f_{q=1}(x)=A^2\cos^2\,k(x-x_0)
\label{fq1}
\end{equation}

\noindent where $k$ is a constant to be
determined below and $A$, $x_0$ are arbitrary integration constants.

It easy to show that (\ref{f}) has, as a first integral,

\begin{equation}
\dot{f}^2+I_q\ f^2+\lambda \ Q\ f^{3-q}=c\ f^{2-q},  \label{primera}
\end{equation}

\noindent where $c$ is an integration constant. This equation involves
Fisher's generalized information for translation families. We must solve it
having (\ref{Iescortq}) in mind. In order to establish the consistency
between (\ref{primera}) and (\ref{Iescortq}) we introduce a set of
normalized variables

\begin{equation}
z=\int \,\sqrt{I_{q}}\,dx,\qquad \bar{\lambda}=\frac{\lambda Q}{I_{q}}%
,\qquad \bar{c}=\frac{c}{I_{q}},
\end{equation}

\noindent (the integral is an indefinite one) in terms of which Eqs. (\ref
{Iescortq}), (\ref{Jq}), (\ref{f}), and (\ref{primera}) are transformed into
\begin{equation}
1=\frac{\int f^{q-2}{f^{\prime }}^{2}dz}{\int f^{q}dz},
\end{equation}

\begin{equation}
J_{q}\,=\bar{\lambda}\,\frac{I_{q}}{\int f^{q}dz}\int \ f(z)\ dz,
\label{Jqn}
\end{equation}
(an indefinite integral),
\begin{equation}
2f\ f^{\prime \prime }+(q-2)\ f^{\prime }{}^{2}+q\ f^{2}+\bar{\lambda}\
f^{3-q}=0,  \label{fn}
\end{equation}
and
\begin{equation}
f^{\prime }{}^{2}+f^{2}+\bar{\lambda}\ f^{3-q}=\bar{c}\ f^{2-q}.  \label{f'}
\end{equation}

\noindent

Inserting (\ref{f'}) into (\ref{Iescortq}) we conclude that the integration
constant acquires the aspect

\begin{equation}
\bar{c}=\frac{2\ Q+\bar{\lambda}}{x_{2}-x_{1}},  \label{con}
\end{equation}

\noindent where $x_2$ and $x_1$ are the integration limits, to be fixed by
the remaining parameters of the theory. A quite interesting point is that
{\it the general solution of (\ref{f'}) can be given in closed form} as

\begin{equation}
\int^{z} dz = z - const. = \pm \int {\frac{f^{\frac{q}{2}-1}}{\sqrt{\bar{c}-%
\bar{\lambda}f-f^{q}}}\,df},  \label{cerrada}
\end{equation}
where the constants $\bar{c},\,\,\,\bar{\lambda}$ must be of such nature
that a real $f$ ensues.


\section{Symmetry properties of the EPI probability distribution}


We start by changing variables in (\ref{fn}) to

\begin{equation}  \label{variable}
u=\frac{f^{\prime}(z)}{f(z)}\ ,
\end{equation}

\noindent and obtaining

\begin{equation}
u^{\prime\prime}+\alpha \ u\ u^{\prime}+\beta \ u^3+\gamma \ u=0\ ,
\label{u}
\end{equation}

\noindent with

\begin{equation}  \label{alpha}
\alpha \ =\ (2\ q-1),\,\,\,   \beta =\frac 12\ q\ (q-1),\,\,\,
\gamma=\beta.
\end{equation}
(A complete study of the properties of equation (\ref{u}) is found in \cite
{chimento}). Further, we effect the transformation

\begin{eqnarray}
f &\rightarrow &1/f,
\end{eqnarray}
so that

\begin{equation}
u \rightarrow -u, \,\,\, u^{\prime} \rightarrow -u^{\prime},
\,\,\, u^{\prime\prime}\rightarrow- u^{\prime\prime}.
\end{equation}

If we require that equation (\ref{u}) be invariant under this
transformation, the parameters $\alpha$, $\beta$ and $\gamma$ must change
according to $\alpha \rightarrow -\alpha $, $\beta\to\beta$ and $%
\gamma\to\gamma$ respectively. This entails that the parameter $q$, that
characterizes the degree of non-extensivity of the system, transform as $%
q\rightarrow 1-q$. A property of this type has been called ``duality" by
Naudts \cite{naudts}, although in his case the relationship is of the form $%
q\rightarrow \frac{1}{q}$ (duality between $q>1$ statistics and $q<1$ one).
In our case, the duality arises between two $q$-values whose sum adds up to
unity.

Introducing now into (\ref{fn}) the new variable

\begin{equation}  \label{h}
h=\frac{1}{f},
\end{equation}
we get

\begin{equation}
2hh''-(q+2)h'^2-qh^2-\bar\lambda h^{q+1}=0,
\end{equation}
which under the substitution $q\to 1-q$, becomes

\begin{equation}
2hh''+(q-3)h'^{2}+(q-1)h^{2}-\bar{\lambda}h^{2-q}=0. \label{hn}
\end{equation}

This equation can be rewritten, if we first define
\begin{equation}
w(q)=(-h'^{2}-h^{2}-\bar{\lambda}h^{2-q}+\bar{c}h^{3-q}),
\end{equation}
as

\begin{equation}  \label{hnn}
2hh''+(q-2)h'^2+qh^2-\bar ch^{3-q}+ w(q)=0,
\end{equation}
where the terms in $w(q)$ correspond to the (transformed) first integral of (%
\ref{fn})

\begin{equation}  \label{fp}
f'^2+f^2+\bar\lambda f^{3-q}=\bar cf^{2-q},
\end{equation}
which under (\ref{h}) becomes
\begin{equation}  \label{hp1}
h'^2+h^2+\bar\lambda h^{2-q}=\bar ch^{3-q}.
\end{equation}

\noindent

As a consequence, $w(q)$ in (\ref{hnn}) vanishes and the equation (\ref{fn}%
), under the transformation (\ref{h}), turns out to retain its form,
changing $q\to 1-q$ and $\bar{c}\to -\bar{\lambda}$. It is convenient at
this point to effect a slight change of notation and denote by $f_q$ the
solution to (\ref{fn}) that obtains when the nonextensivity index is $q$.
The above symmetry argument entails
\begin{equation}
f_{q}(\bar{c},\bar{\lambda})\to \frac{1}{f_{1-q}(-\bar{\lambda},-\bar{c})}.
\label{tf}
\end{equation}

\noindent

Using this symmetry property we can re-obtain the probability distribution
 (\ref{fq1})   for $%
q=1$, i.e., the {\it ordinary, extensive one}, in term of the probability
distribution for $q=0$, that can be easily calculated from (\ref{f'})
\begin{equation}
f^{^{\prime }2}=(\bar{c}-1)\ f^{2}-\bar{\lambda}f^{3},\qquad q=0.
\end{equation}

The solutions are
\begin{equation}
f_{0}(z)=\frac{\bar{c}-1}{\bar{\lambda}}\ \left\{ 1-\tanh ^{2} \frac{%
\sqrt{\bar{c}-1}}{2}\ (z-z_0) \right\} \qquad \bar{c}>1,  \label{f01}
\end{equation}
and
\begin{equation}
f_{0}(z)=\frac{\bar{c}-1}{\bar{\lambda}}\ \left\{ 1+\tan ^{2}\frac{%
\sqrt{1-\bar{c}}}{2}\ (z-z_0) \right\} \qquad \bar{c}<1,  \label{f02}
\end{equation}
where the last solution must be normalized in a finite interval. The
symmetry transformation (\ref{tf}) yields now the general solution for $q=1$
\begin{equation}
f_{1}(\bar{c},\bar{\lambda})\to \frac{1}{f_{0}(-\bar{\lambda},-\bar{c})}.
\label{tf1}
\end{equation}
This is to be compared with the result (\ref{fq1}). We start with (\ref{f02}),
effect the transformation (\ref{tf1}) and reach
\begin{equation}
f_{1}(z)=\frac{\bar{c}}{1+\bar{\lambda}}\,
\cos^2\frac{\sqrt{1+\bar{\lambda}}}{2}\,(z-z_0)
\end{equation}
which, after a little algebra that involves also going back to the
$x$ variable adopts indeed the form (\ref{fq1}) with $
A^{2}=c/\lambda'$ and $
k=\sqrt{\lambda}/2$. A similar analysis can be
performed for (\ref{f01}).

We have thus found the general solution for the (extensive) EPI variational
treatment corresponding to a
 $J$ that entails just normalization of the probability distribution.
 Notice that, within the context of Naudts' effort \cite{naudts}%
, the extensive thermostatistics $q=1$ is self-dual. Instead, according to
the present Fisher framework, the self-dual instance obtains for $q=1/2$.


\section{Conclusions}


We have shown that the EPI principle, used in conjunction with a Fisher measure
constructed with escort distributions that depend upon the Tsallis index $q$,
renders a probability distribution endowed with a remarkable symmetry: a
Naudts'-like duality \cite{naudts}.

Tsallis´ enthusiasts had thought, before the advent of Naudts´ work
\cite{naudts}, that   a different statistics obtains for {\it each different
value of the nonextensivity index $q$}. The duality concept is then important
because it ascribes the same statistics to a given pair of (suitably related)
$q$-values. We have shown here that such a pair can be selected in two distinct
manners, i.e., \`a la Naudts or \`a la Fisher, and have detailed the
prescription corresponding to  the latter choice.

Finally, we have also ascertained which is the general (normalized) probability
distribution that extremizes the physical information.

\end{document}